# Transformative Effect of Oxygen Plasma to Upshot the Structural and Electrical Properties of $Pr_{0.5}Sr_{0.5}MnO_3$ Manganites

Pronita Chettri[a], Bhakta Kunwar[b], Gurukrishna[c], Suraj Mangavati[c], Arun Sarma[d], Ashok Rao[c], C. Devaraja[e], Utpal Deka[e,a] *

[a]Department of Physics, Sikkim Manipal Institute of Technology, Sikkim Manipal University, Majitar, East Sikkim, Sikkim, 737136, India

[b]Department of Physics, Nar Bahadur Bhandari Government College, Tadong, East Sikkim, 737102, India

[c]Department of Physics, Manipal Institute of Technology, Manipal Academy of Higher Education, Manipal, Karnataka, 576104, India

[d] North East Centre for Technology Application and Reach, Upper Nongrim Hills, Shillong 793003, Meghalaya, India

[e]Department of Physics, Manipal Institute of Technology, Bengaluru, Manipal Academy of Higher Education, Manipal, Karnataka, 576104, India

**Corresponding author:** deka.utpal@manipal.edu

**Abstract**

A methodical inquiry of the outcome of oxygen plasma exposure in low bandwidth compounds belonging to the perovskite family $Pr_{1-x}Sr_xMnO_3$ manganites where $x = 0.5$, has been presented in this communication by comparing the structural and transport properties of the untreated and plasma treated samples. It is witnessed that the high temperature transmission is carried out by small polarons while the low-temperature transmission is attributed to variable range polarons. The changes in the transport properties may be attributed to the structural modification due to plasma exposure as revealed by the Rietveld analysis of the X-ray diffraction pattern. Further, oxygen plasma exposure boosts the conductivity due to the integration of oxygen ions in the plasma-exposed samples, thereby rendering them oxygen-rich.

## 1. Introduction

Perovskite manganites are highly correlated systems exhibiting diverse phenomena like colossal magnetoresistance (CMR) [1,2], high Curie temperature ($T_C$) [3], colossal thermoelectric power (TEP) [4], piezoelectricity [5], and magnetocaloric effect [6]. The general chemical symbol for such materials is depicted by the formula $REMnO_3$, where RE represents the rare-earth element. Furthermore, the literature reveals that their properties can be modulated through appropriate insertion of dopants at the RE- or Mn-site. Manganites doped at the RE-site are represented as $RE_{1-x}B_xMnO_3$, where B = alkaline earth element, and $x$ = mole fraction of the dopant. Doped manganites in general, and half-doped manganites $RE_{0.5}B_{0.5}MnO_3$ in particular, have attracted considerable attention owing to their exhibition of diverse and fascinating phenomena like ferromagnetic (FM) to charge ordered antiferromagnetic (CO-AFM) state transition [5], spin-glass insulating state [7], memory and chaos-like effects [8,9] and colossal thermopower [4,10]. Numerous studies have been conducted to explore the properties of $RE_{0.5}B_{0.5}MnO_3$, which have indicated that they can be chosen as the likely composites for memory devices, magnetic refrigeration, solid oxide fuel cells, and solar cells [11–14].

Scientific research in material science primarily deals with the synthesis of materials with an aim for practical applications which necessitate the modification of their properties. However, modifications are challenging with conventional processes of synthesis like doping, and thermal annealing. Alternatively, materials are subjected to various types of radiation like ion beams [15–17], electron beams [18,19], neutron beams [20,21], gamma rays [22,23], and X-rays [24] to modify their physical properties. Ion and electron beam radiation have been used extensively to alter the properties of manganites which is largely attributed to the production of defects in the investigated systems [16,25–27] These studies improve the likelihood of improvisation of the properties of manganites by exposing them to plasma, as plasma is an equilibrium mixture consisting of ions and electrons in addition to neutral atoms.

This notion is supported by a study that has reported an increase in oxygen content in $La_{0.7}Ca_{0.3}MnO_{3-\delta}$ due to oxygen plasma exposure [28]. In yet another report, oxygen plasma treatment led to a rise in conductivity for $La_{0.7}Ca_{0.3}MnO_{3-\delta}$ thin film [29]. Further, few recent studies have reported modification of the structural and transport properties of doped manganites through plasma exposure [30–32]. However, additional research is needed to fully comprehend the influence of plasma irradiation on the characteristics of perovskites in macro or bulk forms. The beauty of plasma is that it forms a plasma sheath with an interacting surface that allows the energetic ions to impinge on the surface of the substrate [33,34]. In the case of oxygen plasma, reactive oxygen ions will hit on the surface and penetrate the substrate depending on the plasma power. Thus, the examination of the outcome of low-pressure oxygen plasma exposure on the structural and transport properties of bulk $Pr_{0.5}Sr_{0.5}MnO_3$(PSMO) manganites is very important which has proven to alter the oxygen content and changing the excited state of the Mn in the manganate. We have focused on the explanation of the structural and carriage properties through X-ray diffraction (XRD), Scanning Electron Microscopy (SEM), temperature corelated electrical resistivity, and Seebeck coefficient measurements.

## 2. Materials and Methods

Polycrystalline samples of half-doped Prasedyomium Strontium Manganites: $Pr_{0.5}Sr_{0.5}MnO_3$ were produced by solid-state reaction technique. High-purity $Pr_6O_{11}$, $MnO_2$, and $SrCO_3$ powders were ground in an agate mortar for around 5-6 h, followed by calcination of the mixture at 1100°C for 24 h. This step was backed by a couple of intermittent grinding and calcination. Finally, a hydraulic press was used to press the powder into rectangular thin pellets. The final step in the process involved the sintering of the pellets at 1250 °C for around 36 h. Post-sintering, the samples were allowed to attain room temperature naturally. Then, they were exposed to oxygen plasma for 1 and 3 minutes. These samples and the unexposed one are hereafter referred to as PSM1, PSM3, and PSM0 respectively. The oxygen plasma was produced through the conventional glow discharge method [35]. Room temperature X-ray diffractogram data (XRD) of all the samples were attained with the help of Rigaku Miniflex diffractometer, where Cu-$K_\alpha$ radiation ($\lambda$=1.54 Å) acts as the X-ray source, in the range of 25°–80° with a step count of about 0.02°. XRD data is applied to verify the phase and crystallinity of the samples. EVO MA18 with Oxford EDS(X-act) was used to carry out EDS measurements at room temperature. This measurement was necessary to confirm the stoichiometry of the prepared samples. Surface scanning electron microscopy (SEM) images of all the samples were taken to investigate the effect of plasma exposure on the surface morphologies. Temperature-dependent resistivity of the PSM0, PSM1, and PSM3 were recorded using the conventional four-probe technique. Further, temperature-dependent thermoelectric power (TEP) was also recorded with the help of the differential DC method in the temperature range of 10-350 K. In addition, near room temperature Hall co-efficient was measured with the help of the Van der Pauw technique using the Ecopia HMS-5500 Hall measurement system.

## 3. Structural Properties

Rietveld refinement plots of XRD data, as shown in Fig 1, are obtained using the FullProf program. XRD refinement confirms the synthesis of the samples as a single-phase perovskite, with no evidence of a secondary phase within the error limits of the experiment. It is observed that PSM0 has an orthorhombic structure with Pbnm symmetry, which is in confirmation with the literature [36]. Both PSM1 and PSM3 have orthorhombic structures with Pbnm symmetry, implying that plasma exposure has no impact on the symmetry of the system. The parameters confirming the structure of the samples as derived from Rietveld analysis are given in Table 1. It was observed that plasma exposure has no significant effect on the values of lattice parameters *a*, *b*, or *c* and cell volume. Furthermore, it was detected that the average bond length Mn-O reduced, while the average bond angle Mn-O-Mn increased on plasma exposure as well as with the time of exposure.

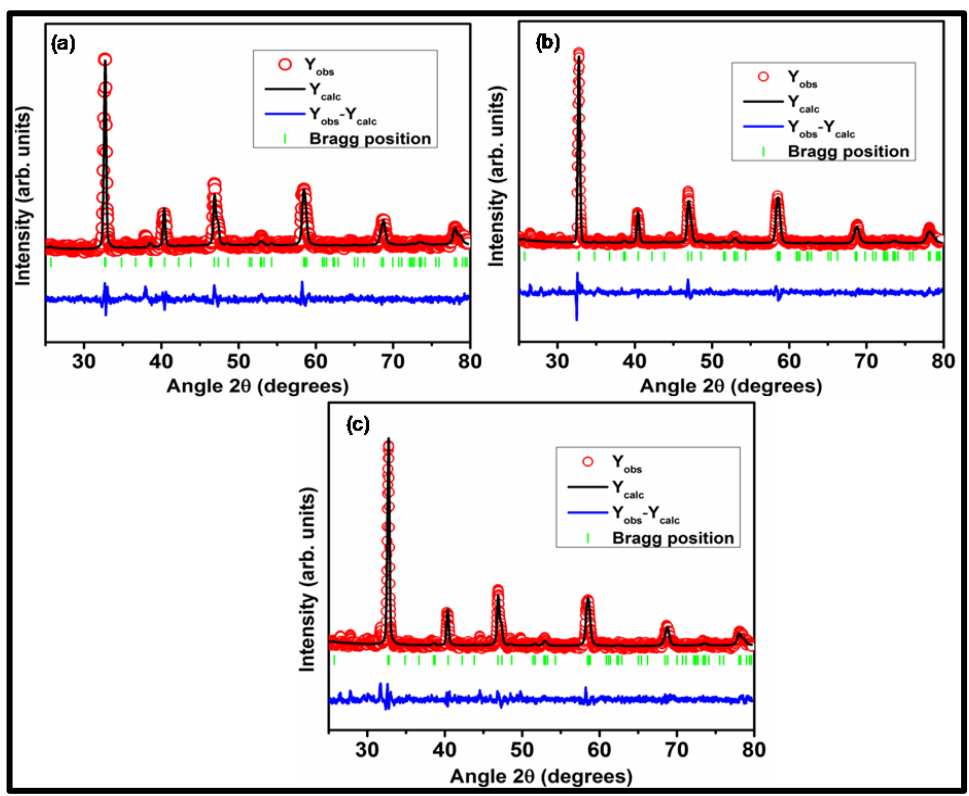

**Figure 1:** Rietveld refinement of XRD data for (a) PSM0 (b) PSM1 and (c) PSM3 samples.

The Williamson-Hall (WH) plot was adopted to determine the crystallite size and strain for PSM0, PSM1, and PSM3 as shown in Fig 3. The plot is fitted with the experimental data through the expression [37]

$$\beta cos\theta = \frac{k\lambda}{D} + 4\varepsilon sin\theta \qquad (1)$$

Reduction in the crystallite size is detected with an increase in the time of plasma exposure which is also supported by SEM images. In addition, the slope of the WH plot determines the strain of the samples, which also depicted a decline with plasma exposure. Oxygen plasma is an ensemble of oxygen radicals, energetic photons, and charged species that are plausibly accountable for the reduction of particle size [38]. Thus, it is indicative that plasma can be employed for the reduction of particle size in nanoparticles.

Further, the grain size estimated through SEM images is quite large in comparison to the values attained from XRD analysis. This variance is accredited to the fact that SEM analysis gives the size of secondary particles while XRD analysis that of primary particles. The SEM micrographs in Fig 3 reveal a homogeneous manganite phase with uniform granular morphology for PSM0 and PSM1. However, the grains are smaller in size for the latter. Interestingly, single crystals are seemingly dispersed in the polycrystalline matrix of PSM3. This may be attributed to the possible local heating of the samples due to exposure for a longer duration leading to the breaking of the particles and ultimately fusing them. Furthermore, according to EDX analysis, no substantial stoichiometric loss of elements for the PSM0 sample is detected within experimental limits. Thus, the obtained compositions of the elements are observed to be close to the nominal one as shown in Table 2.

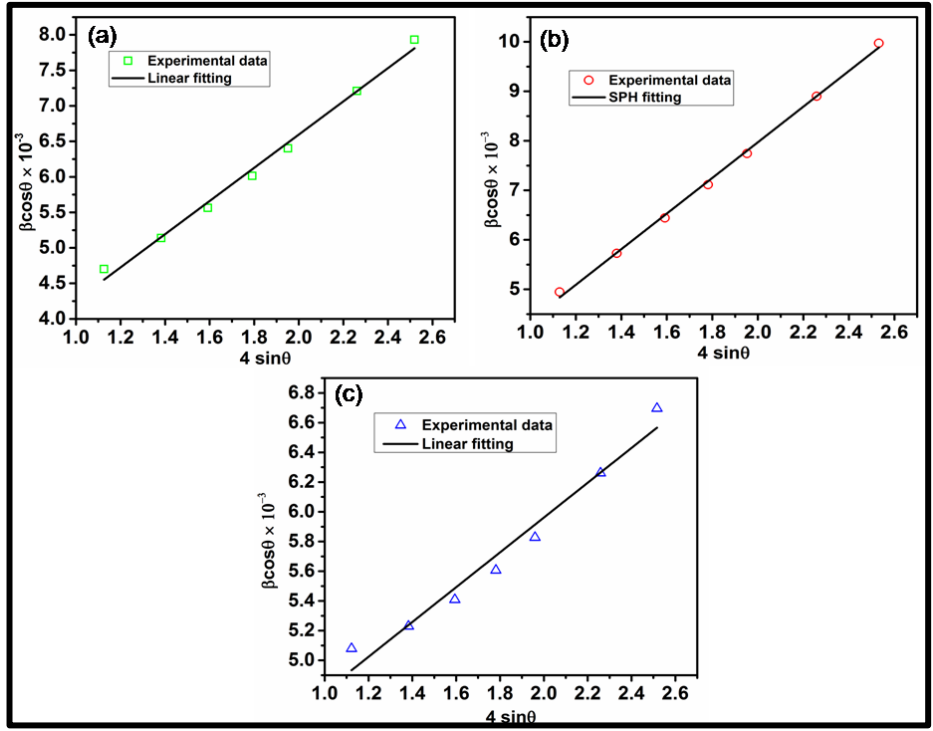

**Figure 2:** Williamson Hall plot for (a) PSM0 (b) PSM1 and (c) PSM3 samples

The electronic bandwidth *W* is evaluated from the XRD analysis with the help of the empirical formula [39]

$$W \propto \frac{cos\{(\pi - <Mn-O-Mn>)/2\}}{d_{Mn-O}^{3.5}} \qquad (2)$$

The transportation of electrons in manganites is subjected to the separation and angles between neighboring atoms. As a consequence, the electron hopping is regulated by *W*. It is observed that by increasing the duration of plasma irradiation the bandwidth also gets amplified [40], thereby resulting in enhancement of the conductivity. This is in confirmation of the results of electrical resistivity study as described in Sec 4.1.

**Table 1:** Structural parameters of PSMO manganites derived from Rietveld refinement of XRD pattern

| Sample | PSM0 | PSM1 | PSM3 |
|---|---|---|---|
| **a (Å)** | 5.484 (0.001) | 5.476 (0.002) | 5.488 (0.001) |
| **b (Å)** | 5.461 (0.002) | 5.473 (0.003) | 5.467 (0.001) |

| c (Å) | 7.674 (0.002) | 7.690 (0.002) | 7.686 (0.001) |
|---|---|---|---|
| Cell volume (Å$^3$) | 229.83 (0.10) | 230.51 (0.15) | 230.70 (0.08) |
| Mn-O1 (Å) | 1.999 | 1.970 | 1.960 |
| Mn-O2 (Å) | 1.958 | 1.969 | 1.975 |
| Average Mn-O (Å) | 1.979 | 1.969 | 1.968 |
| Mn-O1-Mn (degrees) | 147.18 | 154.70 | 157.10 |
| Mn-O2-Mn (degrees) | 162.04 | 158.66 | 157.36 |
| Average Mn-O-Mn (degrees) | 154.61 | 156.68 | 157.21 |
| W (Å$^{-3.5}$) | 0.089 | 0.091 | 0.092 |
| Crystallite size (nm) from WH plot | 32.26 | 23.47 | 21.54 |
| Strain from WH method ($\times 10^{-2}$) | 1.02 | 0.59 | 0.34 |
| Grain size (nm) from SEM | 140.60 | 72.34 | 64.56 |
| $\chi^2$ | 1.32 | 1.48 | 0.98 |

**Table 2:** Elemental compositional analysis of PSM0 sample using EDS

| $x$ | Element | Nominal composition | Experimental composition |
|---|---|---|---|
| **0.5** | Pr | 0.50 | 0.508 |
| | Sr | 0.50 | 0.492 |
| | Mn | 1.00 | 0.98 |

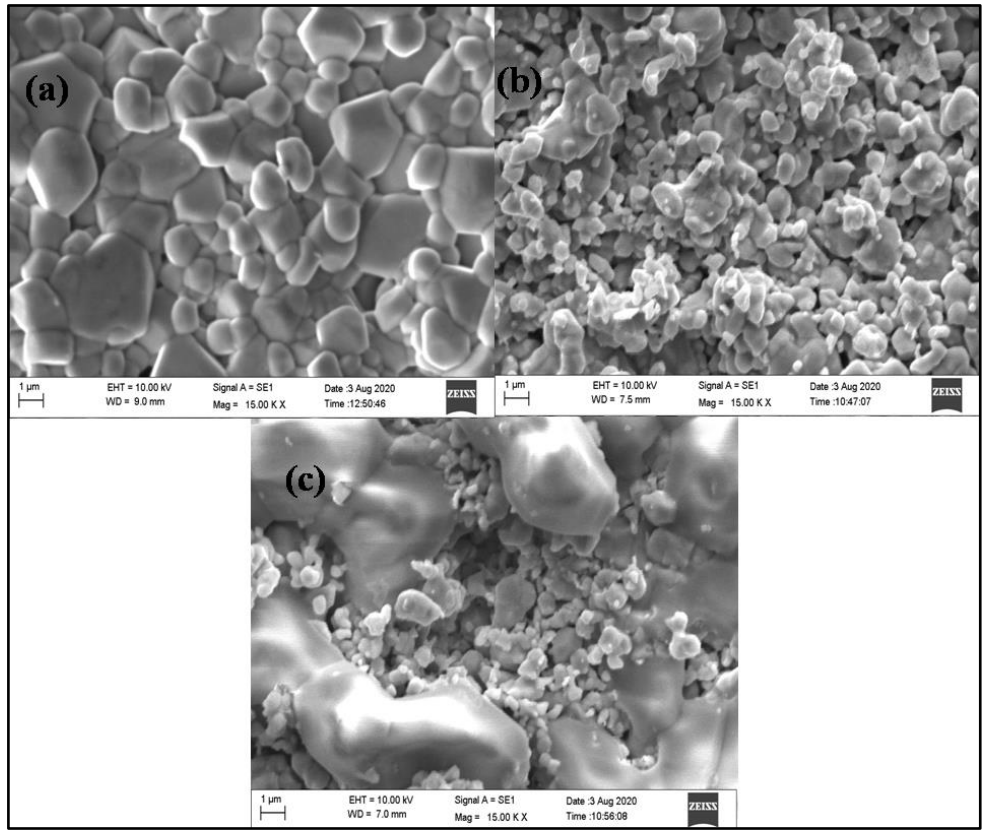

**Figure 3:** SEM micrographs for (a) PSM0 (b) PSM1 and (c) PSM3 samples.

## 4. Transport properties

### 4.1. Electrical Resistivity

Temperature-dependent electrical resistivity $\rho(T)$ experimentation was conducted in the temperature range of 80-350 K for all the specimens. The variation of $\rho(T)$ with $T$ is shown in Fig 4 which depicts the insulating nature of all the samples. However, it is observed that on increasing the exposure time $\rho(T)$ decreases as was reported in another study [35].

The enhancement in the conductivity of the plasma-exposed samples can be associated with the gain in the electronic bandwidth due to plasma treatment. As a consequence of plasma treatment, the electronic bandwidth increases which implies greater overlapping of Mn '$e_g$' and O'$2p$' orbitals. This in turn enhances the electron hopping probability between $Mn^{3+}$ and $Mn^{4+}$ ions [41], thereby increasing the conductivity of PSM1 and PSM3 samples. In a previous study, it has been exhibited that oxygen plasma treatment renders the sample to be oxygen-rich, thereby boosting the conductivity [35]. Hence, it may be speculated that absorption of oxygen on plasma exposure may similarly have improved the conductivity in this case too.

The resistivity data is analyzed in two different temperature zones: the high temperature above $\theta_D/2$ and the low-temperature zone below $\theta_D/2$, where $\theta_D$ represents the Debye temperature. The linear fitting of the small polaron hopping (SPH) model with the experimental data deviates at 130.54, 129.70, and 126.90 K for PSM0, PSM1, and PSM3 samples respectively which earmarks the values of $\theta_D/2$ [42] as seen in Fig 5. High-temperature resistivity is observed to follow SPH conduction [43]. Alternatively, the variable range polaron hopping (VRH) model is befitting to analyze low-temperature resistivity [44].

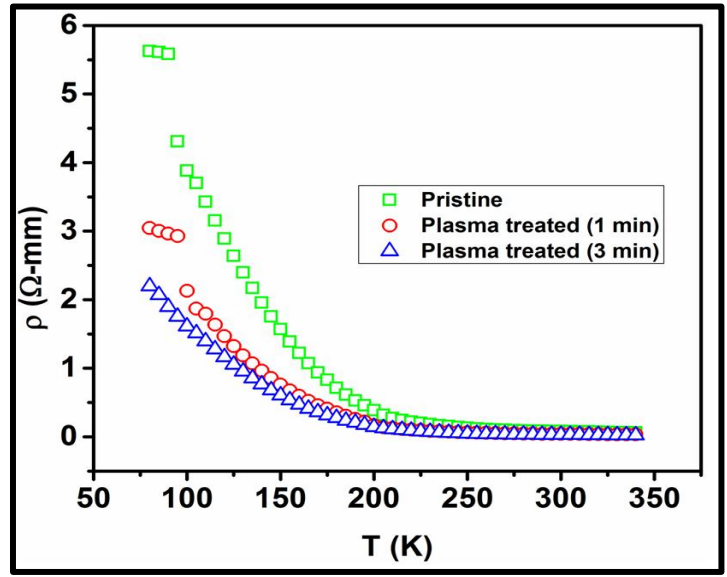

**Figure 4:** **Temperature dependent resistivity variation of pristine and plasma exposed PSMO samples**

The mathematical representation of the adiabatic SPH model is given by equation [43]

$$\rho = \rho_\infty T exp\left(\frac{E_A}{k_B T}\right) \quad (3)$$

where $\rho_\infty$ = residual resistivity, $T$ = absolute temperature, $E_A$ = activation energy and $k_B$ = Boltzmann constant.

In Fig 5 the graph of ln ($\rho/T$) versus $1/T$ is plotted. The close agreement between the theoretical and experimental plots confirms the SPH conduction technique in the temperature range $T \geq \theta_D/2$.

The activation energy $E_A$ was established using the slope of the fitting curve and is tabulated in Table 3. Plasma exposure is observed to decrease $E_A$.

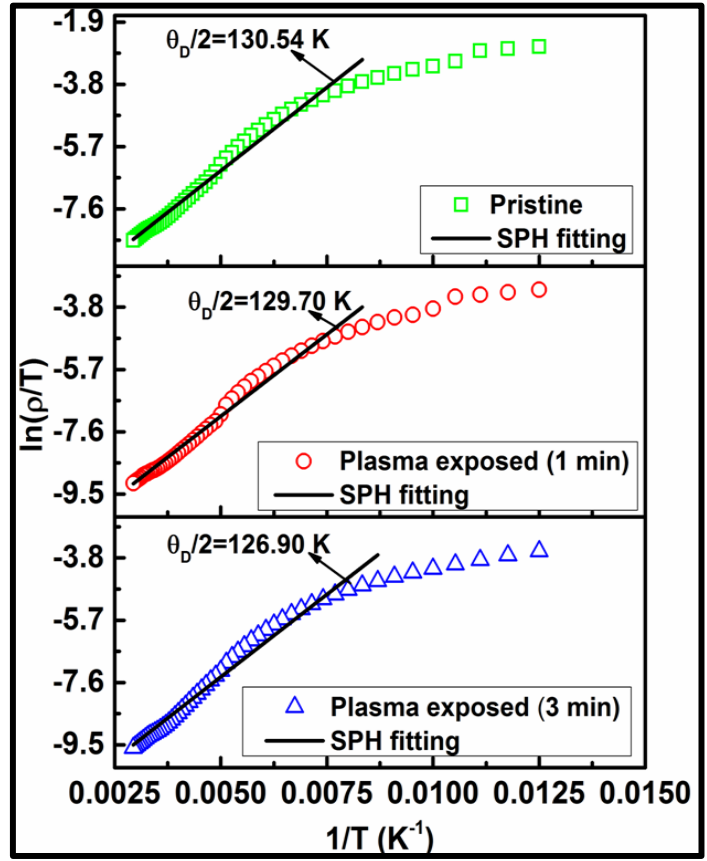

**Figure 5: SPH fitting of resistivity of pristine and plasma exposed PSMO samples**

For the low-temperature resistivity in the range $T < \theta_D/2$, the VRH model is used, which is given by the equation [44]

$$\rho = \rho_0 exp\left(\frac{T_0}{T}\right)^{1/4} \quad (3)$$

where $T_0$ is the characteristic temperature expressed as

$$T_0 = \frac{18\alpha^3}{k_B N(E_F)} \quad (4)$$

and $N(E_F)$ = the density of states at the Fermi level, $\alpha$ = electron wave function decay constant (2.22 nm$^{-1}$) [45].

The experimental data conforms to the VRH model as depicted in Fig 7. The density of states at the Fermi level $N(E_F)$ is calculated by using the value of $T_0$ which is obtained from the slope of this curve. Plasma exposure assists in the increase of the $N(E_F)$ value. This result suggests that the number of charge carriers at the Fermi level increases which in turn augments the conductivity.

The negative Hall coefficient of all the samples was detected through Hall measurement conducted at 325K, which revealed the n-type conductivity in the samples.

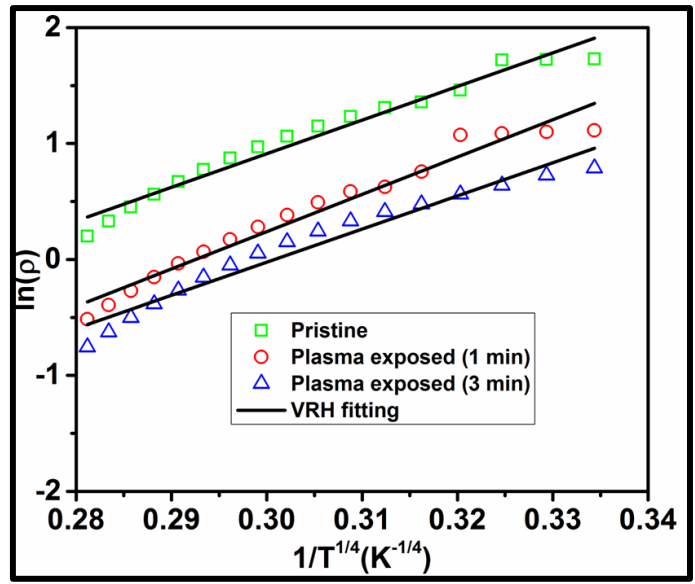

**Figure 6: VRH fitting resistivity of pristine and plasma exposed PSMO samples**

### 4.2. Thermoelectric power

Fig. 7 shows the temperature-related behavior of thermoelectric power (TEP) $S$ for all the specimens in the range 5-300 K. S is negative for all the samples, PSM0, PSM1, and PSM3, for the whole temperature range. This fact suggests that electrons are the predominant charge carriers for these samples. An increase in the absolute value of $S$ ($|S|$) with a rise in temperature is observed for PSM0. $|S|$ decreases as temperature increases for low temperatures (up to ~ 58 K) for PSM1. As the temperature increases further, $|S|$ increases. However, PSM3 shows a similar variation for $S$ with temperature as the pristine sample. Irrespective of the variation shown, PSM1 and PSM3 have $|S|$ to be more than that of PSM0 at all temperatures.

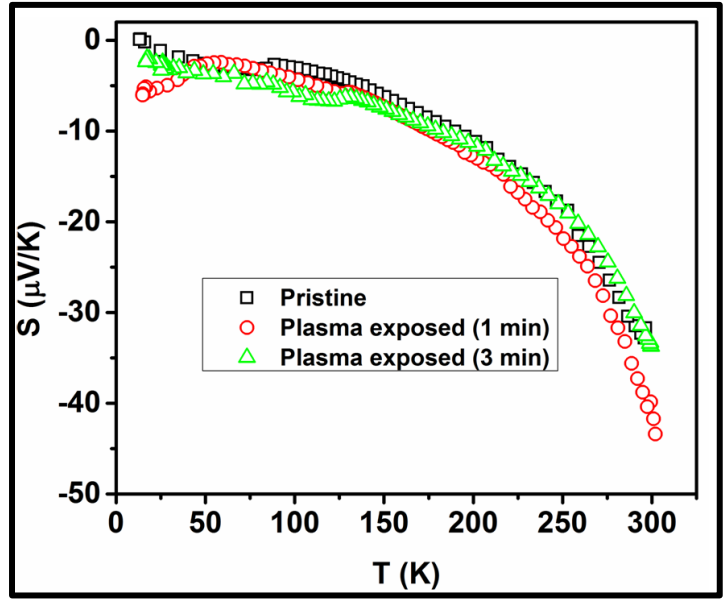

**Figure 7:** Temperature dependent thermopower variation of pristine and plasma exposed PSMO samples

Insight about the process of the conduction is put forward by the analysis of thermoelectric data in addition to that of resistivity data. Therefore, thermoelectric data is also scrutinized using SPH and VRH models. At high temperatures, SPH analysis is done, which is governed by equation [43]

$$S = -\frac{k_B}{e}\left(\frac{\Delta}{k_B T} + \alpha\right) \qquad (5)$$

where $k_B$ = Boltzmann constant, $\Delta$ = activation energy, and $\alpha$ = constant which is connected to the entropy of the charge carriers [46,47].

The fitting for the thermopower data with the SPH model is depicted in Fig 8 which depicts a decent match between the experimental and theoretical data. The slope and intercept of the fitting line are used to determine the activation energy $\Delta$ and $\alpha$ respectively. We obtain $\alpha < 1$ and $E_A \gg \Delta$, thereby indicating the active participation of small polarons in the transport properties of the pristine as well as plasma irradiated samples [48]. It may be noted that a certain amount of the activation energy $E_A$ is utilized for the formation of the carriers and the remaining i.e., $\Delta$ is the activation energy of hopping [49]. It is also observed that the activation energy $\Delta$ increases with plasma irradiation that revitalizes the conductivity in PE samples.

Mathematically, $E_A = \Delta + W_H$, where $W_H$ signifies the minimum energy necessary for the creation of carriers. Further, it may be noted that $W_H = E_b/2$ [49], where $E_b$ is the polaron binding energy. These values decrease with plasma exposure as depicted in Table 3. The variation in $E_b$ is seen to be directly linked to the reduction in average Mn-O bond length which in turn varies with $Mn^{3+}$ content in the samples [50]. This result further supports our speculation of the formation of an oxygen-rich sample due to plasma exposure.

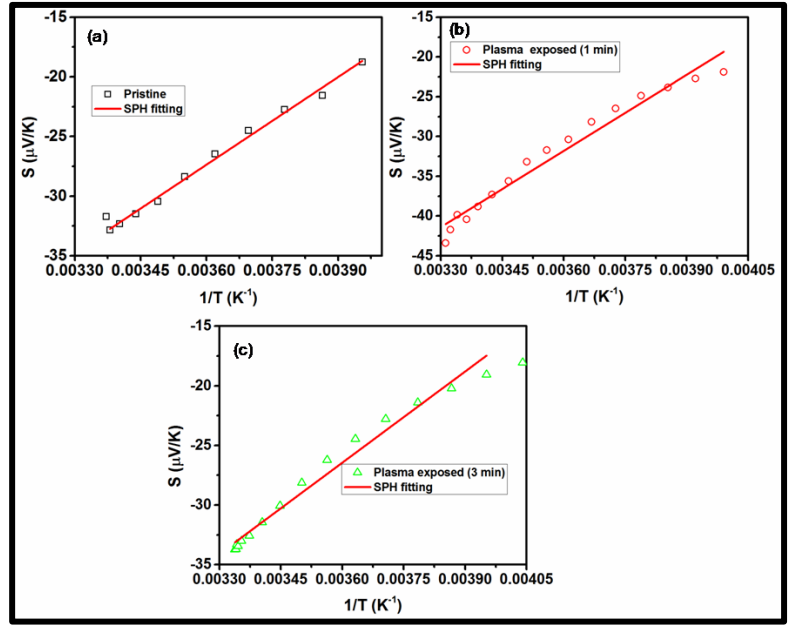

**Figure 8:** **SPH fitting of thermopower variation for (a) PSM0 (b) PSM1 and (c) PSM3 samples**

At low temperatures, the VRH model is used to analyze the thermoelectric power [51]. According to the VRH model, the TEP is defined using

$$S = S_0 + KT^{1/2} \quad (6)$$

$$K = \left(\frac{k_B}{3e}\right)\xi^2 T_0^{1/2}\left[\frac{d}{dE_F}lnN(E_F)\right] \quad (7)$$

where ξ is a factor that governs the width of the energy layer inside which the states are operative in conduction [51]. $N(E_F)$ and $T_0$ have their usual meaning for resistivity. Fig 9 reveals that the VRH model fits well with the experimental data. Thus, we can conclude that TEP at low temperatures adheres to the VRH model. The fitting parameters are depicted in Table 3.

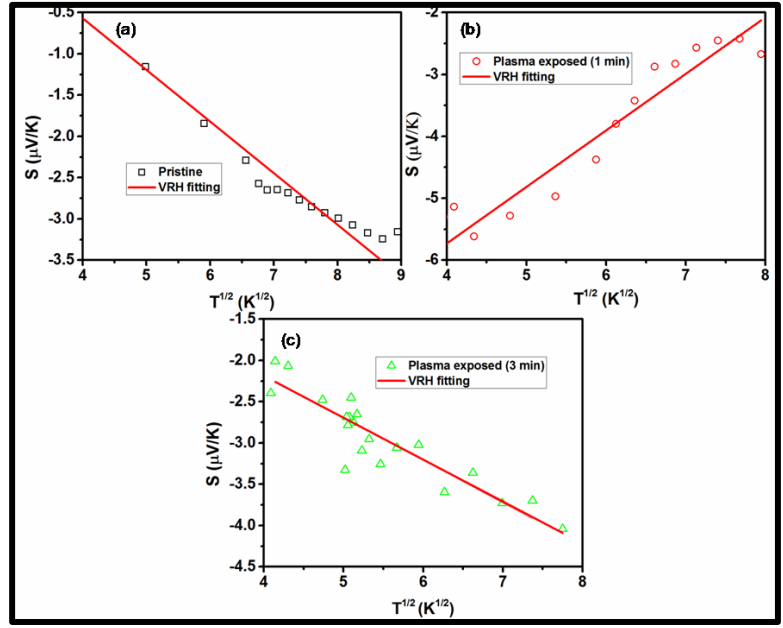

**Figure 9:** VRH fitting of thermopower variation for (a) PSM0 (b) PSM1 and (c) PSM3 samples

**Table 3:** Fitting parameters obtained from resistivity and thermoelectric power data for pristine and plasma exposed PSMO samples

| Sample | SPH Model fitting for resistivity | | VRH Model fitting for resistivity | | SPH Model fitting for thermopower | | VRH Model fitting for thermopower | | Polaron binding energy $E_B$ (meV) |
|---|---|---|---|---|---|---|---|---|---|
| | $E_A$ (meV) | $\rho_\infty \times 10^{-6}$ (Ω-mm) | $T_0 \times 10^5$ (K) | $N(E_F) \times 10^{27}$ (eV$^{-1}$ m$^{-3}$) | $\alpha$ | $\Delta$ (meV) | $S_0$ (μV/K) | $K$ (μV/K$^{1/2}$) | |
| PSM0 | 87.56 | 9.97 | 4.06 | 3.42 | -0.77 | 11.56 | 2.41 | -0.68 | 38.00 |
| PSM1 | 86.07 | 5.44 | 6.68 | 5.62 | -0.86 | 12.81 | -0.59 | -0.48 | 36.63 |
| PSM3 | 86.70 | 3.88 | 3.20 | 7.12 | -0.39 | 25.48 | -0.51 | -0.52 | 30.61 |

## 5. Conclusions

Investigation of the upshot of plasma on the structural and transmission properties of PSMO manganites has been the focus of this work. Room temperature XRD was used to assess the phase and purity of pristine and plasma-exposed samples. XRD data reveal the existence of a single phase for all the samples. Rietveld refinement of the XRD data is used to establish their structural characteristics. All the samples are orthorhombic structures with Pbnm symmetry. Plasma exposure has a considerable effect on Mn-O bond lengths and Mn-O-Mn bond angles. On the contrary, it does not seem to induce any radical change in the attributes of lattice parameters *a*, *b,* or *c* and the cell volume. This results in an increase in the electronic bandwidth. The crystallite size is reduced as a result of the plasma treatment.

The pristine PSMO sample exhibits insulating behavior at all temperatures. This trend is not altered even when the samples are exposed to oxygen plasma. SPH analysis of the resistivity data acclaims that small polarons are accountable for the transport mechanism at high temperatures. Additionally, at low temperatures, variable range polarons are in control for conduction. These actualities are also supported by the analysis of TEP data. TEP investigation also disclose *n*-type conductivity in the manganites under study. Plasma exposure leads to a diminution in the absolute value of *S* at all temperatures for pristine and PE (for 3 min) samples. It is understood that the absolute value of *S* reduces up to T ~ 58 K for PSM1 and thereafter rises with rise in temperature. However, the absolute value of *S* increases for both PSM1 and PSM3. Overall, oxygen plasma played a noteworthy transformation in the overall properties of manganites. It will be stimulating to understand if such a characteristic is universal for all different types of perovskite manganites.

**Acknowledgments**

This work was supported by the Department of Atomic Energy, Board of Research in Nuclear Sciences (DAE-BRNS), Government of India, and Manipal Academy for Higher Learning, Karnataka, under Research Mobility Grant.